\documentclass[12pt]{article}
\pdfoutput =1
\usepackage{float} 
\usepackage{amsmath}
\usepackage{amsfonts}   
\usepackage{amssymb}

\begin{document}
\date{}
\title{{\bf{\Large Holographic charge diffusion in non relativistic branes}}}
\author{
 {\bf {\normalsize Dibakar Roychowdhury}$
$\thanks{E-mail:  dibakarphys@gmail.com, dibakar@cts.iisc.ernet.in}}\\
 {\normalsize Centre for High Energy Physics, Indian Institute of Science, }
\\{\normalsize C.V. Raman Avenue, Bangalore 560012, Karnataka, India}
}

\maketitle
\begin{abstract}
In this paper, based on the principles of gauge/gravity duality and considering the so called \textit{hydrodynamic} limit we compute various charge transport properties for a class of strongly coupled non relativistic CFTs corresponding to $ z=2 $ fixed point whose dual gravitational counter part could be realized as the consistent truncation of certain non relativistic $ Dp $ branes in the non extremal limit. From our analysis we note that unlike the case for the AdS black branes, the charge diffusion constant in the non relativistic background scales differently with the temperature. This shows a possible violation of the universal bound on the charge conductivity to susceptibility ratio in the context of non relativistic holography. 
 \end{abstract}

\section{Overview and Motivation}
For the past several years, the quantum critical theories with anisotropic scaling namely,
\begin{eqnarray}
t\rightarrow \lambda^{z}t,~~\textbf{x}\rightarrow \lambda \textbf{x}\label{A}
\end{eqnarray}
has received renewed attention due to their remarkable connection with the holographic principle \cite{Taylor:2008tg}. The anisotropic scaling of the above type (\ref{A}) could in principle be realized in various condensed matter applications, for example in the so called \textit{cold atom }systems which are the perfect experimental realizations of fermions at unitarity. These are the systems where it is indeed difficult to carry out perturbative calculations and therefore difficult to solve them analytically. Non relativistic CFTs (with the scale invariance of the above type (\ref{A})) corresponding to the $ z=2 $ fixed point exhibit the so called Schr{\"o}dinger symmetry whose isometry group consists of the generators corresponding to time translation ($ H $), spatial rotations ($ M_{ij} $), spatial translations ($ P_i $), Galilean boosts ($ K_i $), special conformal transformation ($ C $), dilatation ($ D $) and the number operator ($ N $).

It was realized for the first time in \cite{Son:2008ye}-\cite{Balasubramanian:2008dm} that the Schr{\"o}dinger isometry group in $ D+1 $ dimensions (which is basically the symmetry group corresponding to fermions at unitariry for $ D=2 $) could in principle be realized holographically as a dual of $D+3$ dimensional geometry where the additional dimension was required in order to understand the Galilean boosts, special conformal transformation and the mass operator. Saying it in other way, the additional two directions could be thought of as the light cone directions where one of the light cone coordinates plays the role of time and the other direction turns out to be a null circle with certain periodicity that could be related to the particle number of the dual non relativistic CFT at strong coupling. The holographic model that was originally proposed in \cite{Son:2008ye}-\cite{Balasubramanian:2008dm} had been found to describe the large $ N $ limit of the so called non relativistic CFTs at zero temperature as well as at zero chemical potential ($ \mu $). These solutions had been found to have their natural embedding into type IIB string theory and were generated from the extremal $ D3 $ branes by applying so called Null Melvin Twist \cite{Adams:2008wt}.  Later on this construction has been further extended for systems with finite temperature as well as chemical potential with a number of possible applications in different directions \cite{Adams:2008wt}-\cite{Wang:2013tv}. The underlying principle behind these holographic constructions rests on the fact that there must be some original string theory description which in the low energy limit produces some effective gravitational description whose isometry group in stead of being the usual Poincare group is the Schr{\"o}dinger group which is eventually the symmetry group of fermions at unitarity. 

In the context of AdS/CFT duality it is the celebrated fact that the hydrodynamic description of large $ N $ CFTs could be best described in terms of a dual gravitational theory embedded in an Anti-de Sitter (AdS) space in the presence of black holes where using the so called holographic principles on can in fact calculate the various transport coefficients of the fluid at strong coupling. Surprisingly it has been observed that for a large class of CFTs (that admit dual gravitational description embedded into an AdS space) these transport coefficients could be uniquely fixed in terms of thermodynamics and in particular it has been observed that for $ D\geq 2+1 $ the so called Einstein's relation namely the ratio between the DC electrical conductivity ($ \sigma_{DC} $) to charge susceptibility ($ \chi $) is universal \cite{Kovtun:2008kx}-\cite{Herzog:2002fn},
\begin{eqnarray}
\mathcal{D}=\frac{\sigma_{DC}}{\chi}=\frac{1}{4 \pi T}\frac{D}{D-2}.\label{Sigma}
\end{eqnarray}       
The above universality (\ref{Sigma}) could be thought of as an artefact of the universal connection between the hydrodynamic transport coefficients (that describes the physics at large distances) to that with the central charges (that describes the physics at short distances) for a large class of CFTs in more than two dimensions.

Keeping the spirit of these analysis and considering the so called hydrodynamic limit the purpose of the present article is to explore the above universality bound for finite temperature non relativistic (large $ N $) CFTs corresponding to $ z=2 $ fixed point whose dual gravitational counterpart in the non extremal limit could be realized as a consistent truncation of non relativistic $ Dp $ baranes that may or may not preserve the hyperscaling.

In our analysis we consider the hyperscaling preserving non extremal Reissner-Nordstr{\"o}m-Schr{\"o}dinger black branes \cite{Adams:2009dm}-\cite{Cremonesi:2009gy} ($ RNSch_5 $) in ($ 4+1 $) dimensions and compute the above ratio (\ref{Sigma}) over that background. In order to address this issue, what one essentially needs to compute is the retarded Minkowski correlators for $ RNSch_5 $ black branes following the standard prescription of \cite{Son:2002sd}-\cite{Policastro:2002se}. This in fact enables us to extract the DC conductivity for such systems and to check the Einstein's relation for the first time in the context of non relativistic branes\footnote{In the context of Lifshitz black branes the DC conductivity as well as the Einstein's relation has been studied in \cite{Pang:2009wa}-\cite{Lemos:2011gy}}. 

To summarize the key findings of our analysis, we note that the diffusion constant ($ \mathcal{D}=\sigma_{DC}/\chi $) for non extremal $ RNSch_5 $ black branes eventually shows a different scaling with the temperature namely,
\begin{eqnarray}
\mathcal{D} \sim T^{-3}\label{NE}.
\end{eqnarray}
 A few comments are to be made at this stage regarding the above result in (\ref{NE}). First of all, one should take a note on the fact that like the $ \eta/s $ ratio, the $ \sigma_{DC}/\chi $ ratio (for a certain class of non relativistic CFTs in the large t'Hooft coupling limit) also does not depend on the powers of $ N $ which is in spirit is quite similar to what is observed for the pure AdS black branes. On the other hand, this result is dramatically different from that of the usual bound as mentioned in (\ref{Sigma}). This observation therefore clearly confirms the non universality of the ratio (\ref{Sigma}) in the context of non relativistic branes. Such a deviation from the universality is the consequence of the fact that in the  non relativistic limit there is no upper bound for the speed of light ($ \upsilon \rightarrow \infty $) and therefore the above bound (\ref{Sigma}) is expected to be violated in the non relativistic scenario \cite{Kovtun:2008kx}.

The organisation of the paper is the following: In Section 2, we briefly discuss the basic holographic set up in order to describe the dual non relativistic hyperscaling preserving CFTs corresponding to $ z=2 $ fixed point. In Section 3, following the prescription of \cite{Son:2002sd}-\cite{Policastro:2002se} we compute the charge diffusion coefficient ($ \mathcal{D} $) for hyperscaling preserving CFTs at strong coupling. Finally, we conclude in Section 4.

\section{The model}
Before we actually start our analysis, it is always good to have a brief overview of the background over which the analysis is performed. The goal of the present discussion is therefore to provide a very formal introduction to the holographic set up in the bulk namely the hyperscaling preserving charged Schr{\"o}dinger black brane \cite{Cremonesi:2009gy}, that essentially corresponds to some equilibrium configuration of the dual non relativistic CFT in the \textit{non extremal} limit i.e; at finite temperature ($ T $) and charge density ($ \varrho $). 

In our analysis we mostly follow the prescription of \cite{Son:2002sd}-\cite{Policastro:2002se}, where we introduce an additional $ U(1) $ field strength two form\footnote{This additional Maxwell field $\mathcal{F}_{(2)} $ could be treated as the source of some real electromagnetic field over the $ RNSch_5 $ background. This in fact is in the same line as that of the earlier analysis \cite{Pang:2009pd}-\cite{Brynjolfsson:2009ct} where apart from having the auxiliary gauge fields (that is generated during Kaluza-Klein reduction) one can have some additional field strength two form that does not affect the geometry of the space time (unlike the auxiliary gauge fields) but on the other hand could produce some real physical phenomenon.} $\mathcal{F}_{(2)} $ namely,
\begin{eqnarray}
S_M = -\frac{1}{4g^{2}_{SG}}\int d^{5}x \sqrt{-g}\mathcal{F}_{ab}\mathcal{F}^{ab}\label{E1}
\end{eqnarray}
 which we treat \textit{perturbatively} over the fixed background of the five dimensional Reissner-Nordstr{\"o}m-Schr{\"o}dinger black brane ($ RNSch_5 $) solution that could be formally expressed in the Einstein frame as\footnote{These solutions (\ref{E2}) in principle could be obtained by following a number of systematic steps. Roughly the procedure is the following: To start with one needs to first uplift the five dimensional $ RN AdS_{5} $ black brane to its ten dimensional type IIB supergravity version and then in the next step Kaluza Klein reduce it back to five dimensions again followed by a Null Melvin Twist.} \cite{Cremonesi:2009gy},
\begin{eqnarray}
ds_{5}^{2}&=&\frac{\mathcal{K}^{1/3}L^{2}}{u^{2}}\left[ - \frac{f}{\mathcal{K}}d\tau^{2}-\frac{f \beta^{2}L^{4}}{u^{2}\mathcal{K}}(d\tau + dy)^{2}+ \frac{dy^{2}}{\mathcal{K}}  + \frac{du^{2}}{f}+dx^{2}+dz^{2}  \right] \nonumber\\
f(u)&=&1-(1+\mathcal{Q}^{2})\left(\frac{u}{u_{H}} \right)^{4}+\mathcal{Q}^{2}\left(\frac{u}{u_{H}} \right)^{6},~~\mathcal{K}(u)= 1+\frac{\beta^{2}L^{4}u^{2}}{u_H^{4}}\left[ 1+\mathcal{Q}^{2}\left( 1-\left(\frac{u}{u_H} \right) ^{2}\right) \right] \nonumber\\
A&=&\Phi d\tau ,~~\Phi =\mu\left[ 1-\left(\frac{u}{u_H} \right) ^{2}\right] ,~~\mathcal{Q}=\frac{2\mu u_H}{3}.\label{E2}  
\end{eqnarray}

Note that the above metric (\ref{E2}) has been expressed in the so called Poincare chart with $ u $ as the radial coordinate in that chart such that the horizon is placed at $ u=u_H $ and the asymptotic boundary is located at $ u=0 $. Moreover here $ A $ is the massless gauge field that was originally present in the dimensionally reduced version of the gauge super gravity in five dimensions and $ \mu $ is the corresponding chemical potential\footnote{Apart from these fields there also exist a massive gauge field ($ A_{M} $) and a non trivial dilaton ($ \phi $) that appear during the process of the Null Melvin Twist to the original type IIB supergravity theory in ten dimensions \cite{Adams:2009dm}.}. 
The entity $ \mathcal{Q} $ is the $ U(1) $ charge and $ \beta $ is the dimension full parameter ($ [\beta] =\frac{1}{L} $) characterizing the null Melvin twist to the original type IIB supergravity theory in ten dimensions and which is also associated with the particle number in the dual theory. 

Following the prescription of \cite{Adams:2009dm}, the Hawking temperature corresponding to the $ RNSch_5 $ black brane turns out to be,
\begin{eqnarray}
T&=&\left[\frac{1}{2 \pi \sqrt{g_{uu}}} \partial_u\sqrt{\mathcal{G}_{\tau\tau}}\right]_{u=u_H} \equiv \frac{|2-\mathcal{Q}^{2}|}{2\pi u_H}\nonumber\\
\mathcal{G}_{\tau\tau}&=& -\frac{L^{2}f}{u^{2}\mathcal{K}^{2/3}}.
 \label{E3} 
\end{eqnarray}

The interesting observation that one should make at this stage is that the Hawking temperature of the black brane does not contain the parameter $ \beta $ explicitly and therefore it does not get affected during the entire Melvinization procedure. In other words this is the same Hawking temperature as that of the pre Melvinized solution \cite{Gubser:2009qm}. This is the consequence of the fact that the Killing generator of the horizon essentially remains the same both before and after the twist which therefore preserves the horizon structure \cite{Adams:2008wt},\cite{Adams:2009dm}.   
Finally, one should also notice from (\ref{E3}) that the temperature ($ T $) has received an inverse dimension of length namely, $ [T]=L^{-1} $. With the above set up in hand, our next task would be to calculate the various charge transport coefficients corresponding to the above non extremal $ RNSch_5 $ black brane solution (\ref{E2}).

\section{Diffusion constant}
Hydrodynamics could be thought of as an effective description of an interacting QFT at long wavelength and/or low frequency limit. In the framework of AdS/CFT duality it is  the celebrated fact that the IR behaviour of an interacting finite temperature QFT could in principle be described in terms of a dual gravitational theory in the bulk that contains certain black brane at some finite Hawking temperature ($ T $). Considering the fluctuations of the $ U(1) $ gauge fields over the fixed background of the black brane one could in fact study the diffusive modes that satisfy certain dispersion relation like $ \omega = -i \mathcal{D}q^{2} $. Here $ \mathcal{D} $ is called the charge diffusion coefficient that confirms the diffusion of conserved $ U(1) $ charges in a hydrodynamic flow \cite{Kovtun:2003wp}.

The aim of this part of the analysis is to explicitly calculate the charge diffusion coefficient ($\mathcal{D}=\sigma_{DC}/\chi $) for the non extremal $ RNSch_5 $ black brane solution given above in (\ref{E2}). In order to compute the diffusion coefficient ($ \mathcal{D} $) we need to know the following two entities independently namely the charge susceptibility ($ \chi $) and the DC conductivity ($ \sigma_{DC} $). In the following we shall first discuss the charge susceptibility ($ \chi $) in the context of non relativistic branes (\ref{E2}) and then in the subsequent section we shall discuss an approach based on the Kubo's formula in order to compute the DC conductivity ($ \sigma_{DC} $).


\subsection{Charge susceptibility}

In order to compute the charge susceptibility ($ \chi $) one first needs to solve the temporal part of the Maxwell's equation over the background (\ref{E2}) and study the asymptotic behaviour of the solution near the boundary of the $ RNSch_5 $ space time.  
The temporal part of the Maxwell's equation could be formally expressed as,
\begin{eqnarray}
\mathcal{A}''_{\tau}(u)+\Re(u)\mathcal{A}'_{\tau}(u)=0\label{E4}
\end{eqnarray}
where the function $ \Re (u) $ could be explicitly written as,
\begin{eqnarray}
\Re (u)&=&\frac{\beta ^2 L^4 u_H^6 \left(-2\mathcal{Q}^2 u^6-2 \left(\mathcal{Q}^2+1\right) u^4 u_H^2+9 u_H^6\right)-3 u^2 u_H^{12}+\mathcal{N}(u)}{3 u \left(\beta ^2 L^4 \left(\mathcal{Q}^2 u^6-\left(\mathcal{Q}^2+1\right) u^4 u_H^2+u_H^6\right)-u^2 u_H^6\right) \left(\beta ^2 L^4 u^2 \left(\mathcal{Q}^2 \left(u_H^2-u^2\right)+u_H^2\right)+u_H^6\right)}\nonumber\\
\mathcal{N}(u)&=&-\beta ^4 L^8 u^2 \left(5 \mathcal{Q}^4 u^8-6 \mathcal{Q}^2 \left(\mathcal{Q}^2+1\right) u^6 u_H^2+\left(\mathcal{Q}^2+1\right)^2 u^4 u_H^4-13 \mathcal{Q}^2 u^2 u_H^6+11 \left(\mathcal{Q}^2+1\right) u_H^8\right).\nonumber\\
\label{E5}
\end{eqnarray}

The immediate next step would be to solve (\ref{E4}) in order to extract the boundary behaviour of the gauge field. Although the exact solution of (\ref{E4}) turns out to be too complicated to achieve, still one can manage to solve it for small values of $ u $. This is perfectly logical in the sense that for our purpose it is indeed sufficient to know the near horizon ($ u \sim u_H $) as well as the boundary behaviour ($ u \sim 0 $) of the solutions in the bulk. In the following we note down the exact solution of the above equation (\ref{E4}) near the boundary namely,
\begin{eqnarray}
\mathcal{A}_{\tau} (u)&=&\mathcal{C}_{2} + \mathcal{C}_{1}\ \Xi(u)\nonumber\\
\Xi (u)&=& \frac{3 \beta ^2 L^4 u_H^4 \left(u^2 \left(\beta ^4 L^8 \left(\mathcal{Q}^2+1\right)+3 u_H^4\right)-3 \beta ^2 L^4 u_H^4\right) \exp \left(\frac{u^2 \left(\beta ^4 L^8 \left(\mathcal{Q}^2+1\right)+3 u_H^4\right)}{3 \beta ^2 L^4 u_H^4}\right)}{2 \left(\beta ^4 L^8 \left(\mathcal{Q}^2+1\right)+3 u_H^4\right)^2}.\nonumber\\
\label{E6}
\end{eqnarray}

Note that here $ \mathcal{C}_{1} $ and $ \mathcal{C}_{2} $ are two arbitrary (dimensionful) coefficients that could be related to the chemical potential ($ \mu $) as well as the charge density ($ \varrho $) of the boundary theory in the limit $ u\rightarrow 0 $. Using (\ref{E6}) it is now in fact quite straightforward to compute the charge density for the boundary theory which turns out to be \cite{Kovtun:2008kx},
\begin{eqnarray}
\varrho = \frac{\delta S^{(OS)}_{M}}{\delta\mathcal{A}_{\tau}}|_{u\rightarrow 0}= \frac{\mathcal{C}_{1}L^{5}\beta^{2}}{g^{2}_{SG}}.\label{E7}
\end{eqnarray}

On the other hand the chemical potential turns out to be\footnote{Note that as it is quite evident from (\ref{E2}) that the chemical potential ($ \mu $) has the dimension of energy ($ [\mu]=L^{-1} $) therefore both the coefficients $ \mathcal{C}_{1} $ and $ \mathcal{C}_{2} $ must also have dimensions namely, $ [\mathcal{C}_{1}]=L^{-5}$ and $[\mathcal{C}_{2}]=L^{-1} $. This also suggests the fact that the coupling of the Maxwell action (\ref{E1}) has the dimension of length namely, $ [g_{SG}^{2}]=L $. This is in fact perfectly consistent with the fact that in general in $ D+1 $ dimensions the gauge coupling constant has the dimension of $ L^{D-3} $ \cite{Kovtun:2008kx}.},
\begin{eqnarray}
\mu = \mathcal{A}_{\tau} (0)=\mathcal{C}_2-\frac{9 \beta ^4 \mathcal{C}_1 L^8 u_H^8}{2 \left(\beta ^4 L^8 \left(\mathcal{Q}^2+1\right)+3 u_H^4\right)^2}.\label{E8}
\end{eqnarray}

Finally, using (\ref{E7}) and (\ref{E8}) the charge susceptibility ($ \chi $) finally turns out to be,
\begin{eqnarray}
\chi = \varrho/\mu = \frac{2L^{5}\beta ^2\mathcal{C}_1  \left(\beta ^4 L^8 \left(\mathcal{Q}^2+1\right)+3u_{H}^{4}\right)^2}{g^{2}_{SG}(2\mathcal{C}_2 \left(\beta ^4 L^8 \left(\mathcal{Q}^2+1\right)+3u_{H}^{4}\right)^2-9 \mathcal{C}_1u_{H}^{8}\beta ^4 L^8)}.\label{E9}
\end{eqnarray}

Two points are to be noted at this stage. Firstly, one should take a note on the fact that unlike the temperature ($ T $) the charge susceptibility ($ \chi $) has been found to be explicitly dependent on the entire Melvinization scheme and Secondly, setting $ \mathcal{Q}=0 $ in (\ref{E9}) one should be able to register the value of the charge susceptibility corresponding to the uncharged Schr{\"o}dinger black brane. Finally, from the dimensional arguments one should be able note that the susceptibility ($ \chi $) has the inverse square dimension of length namely\footnote{This dimensionality of $ \chi $ could be easily checked using the dimensional arguments. Since $ [\mu]=L^{-1} $ (see (\ref{E2})), therefore from (\ref{E8}) it is trivial to see that $ [\mathcal{C}_2]=L^{-1} $ and $ [\mathcal{C}_1]=L^{-5} $. Also $ [g_{SG}^{2}]=L $ and $ [\beta]=L^{-1} $. Substituting all these facts into (\ref{E9}) one obtains the above dimensionality for the susceptibility ($ \chi $).}, $ [\chi]=L^{-2} $, which therefore suggests the fact that it goes with the temperature as $ \chi \sim T^{2} $.

\subsection{DC conductivity}
In order to compute the DC conductivity ($ \sigma_{DC}(\omega =0) $) we shall use the so called Kubo's formula which states that in the hydrodynamic limit the DC conductivity could be estimated by knowing the retarded two point (current-current) correlation function namely\footnote{Note that the notion of electrical conductivity in the framework of AdS/CFT duality is a bit shuttle in the sense that the gauge fields do not actually have any dynamics at the boundary. Therefore by the notion of conductivity what we essentially mean is that the boundary theory is weakly gauged with certain small gauge coupling parameter $ e $ and there also exists a gauge covariant derivative of the form $ D_{\mu}=\partial_{\mu}-ie A_{\mu} $. All these facts essentially suggest that the corresponding global $ U(1) $ current at the boundary should also be replaced as $ J_{\mu}\rightarrow e J_{\mu} $ \cite{Kovtun:2008kx}.},
\begin{eqnarray}
\sigma_{DC} = - e^{2}\lim_{\omega \rightarrow 0}\frac{1}{\omega}Im\ \mathcal{G}^{R}_{xx}(\omega , \textbf{q}=0)
\end{eqnarray}
where $ \mathcal{G}^{R}_{xx}(\omega , \textbf{q}=0) $ is the retarded two point correlator,
\begin{eqnarray}
\mathcal{G}^{R}_{xx}(\omega , \textbf{q}=0) = -i \int d\tau\ d\textbf{x}\ e^{i\omega \tau}\ \Theta (t)\ \langle[J_x (\textbf{x}), J_x (0)]\rangle .
\end{eqnarray}

In the present paper we shall compute the retarded Minkowski correlator by following the standard prescription of \cite{Son:2002sd}-\cite{Policastro:2002se}. In order to proceed further we consider the following Fourier decomposition of the gauge field namely,
\begin{eqnarray}
\mathcal{A}_{a}(u,\tau)=\int \frac{d\omega}{\sqrt{2 \pi}} e^{-i \omega  \tau}f_{a(\omega)}(u)\phi^{(a)}_{0}(\omega);~~ (a=\tau, x)\label{E12}
\end{eqnarray}
where we have chosen the spatial momentum along a particular direction (say $ x $) equal to zero. Moreover here $ \phi^{(a)}_{0}(\omega) $ corresponds to the boundary value of the bulk gauge field and could in principle be determined by the boundary condition namely,
\begin{eqnarray}
\mathcal{A}_{a}(0,\tau)=L\int \frac{d\omega}{\sqrt{2 \pi}} e^{-i \omega  \tau}\phi^{(a)}_{0}(\omega)
\end{eqnarray}
where we have used the fact that the value of the function $ f_{a(\omega)}(u) $ at the boundary is simply $ f_{a(\omega)}(u)|_{u=0}=L $. Finally, substituting (\ref{E12}) into (\ref{E1}) we obtain,
\begin{eqnarray}
S_M = S_\tau + S_{x}
\end{eqnarray}
where $ S_\tau $ and $ S_x $ could be individually expressed as,
\begin{eqnarray}
S_\tau &=& -\frac{V}{2g^{2}_{SG}}\int du\ \int d\omega\ d\omega'\ \delta(\omega + \omega')\ a(u)\ f'_{\tau(\omega)}f'_{\tau(\omega')}\phi^{(\tau)}_{0}(\omega)\phi^{(\tau)}_{0}(\omega')\nonumber\\
S_x &=&-\frac{V}{2g^{2}_{SG}}\int du\ \int d\omega\ d\omega'\ \delta(\omega + \omega')\left[b(u)f'_{x(\omega)}f'_{x(\omega')}- c(u)\omega \omega' f_{x(\omega)}f_{x(\omega')}  \right] \phi^{(x)}_{0}(\omega)\phi^{(x)}_{0}(\omega')\nonumber\\
a(u)&=&\sqrt{-g}g^{uu}g^{\tau \tau};~~b(u)=\sqrt{-g}g^{uu}g^{xx};~~c(u)=\sqrt{-g}g^{xx}g^{\tau \tau}\label{E14}
\end{eqnarray}
where $ V(=\int d^{3}\textbf{x}) $ is the three spatial volume.

In the subsequent analysis we shall focus on the part $ S_x $ as we are interested in computing the current-current correlator of the form $ \sim \langle J_x(\textbf{x})J_x (0)\rangle $. The equation of motion corresponding to $f_{x(\omega)} $ turns out to be,
\begin{eqnarray}
f''_{x(\omega)}+\frac{b'(u)}{b(u)}f'_{x(\omega)}-\omega^{2}\frac{c(u)}{b(u)}f_{x(\omega)}=0.\label{E15}
\end{eqnarray}

Substituting (\ref{E15}) into (\ref{E14}) the onshell action corresponding to $ S_x $ turns out to be,
\begin{eqnarray}
S_x^{(os)}= \int d\omega\ \phi^{(x)}_{0}(-\omega)\mathcal{F}(u,\omega)\phi^{(x)}_{0}(\omega)
\end{eqnarray}
where the function $ \mathcal{F}(u,\omega) $ could be formally expressed as,
\begin{eqnarray}
\mathcal{F}(u,\omega) =-\frac{V}{2g^{2}_{SG}} b(u)f_{x(-\omega)}f'_{x(\omega)}\label{E17}
\end{eqnarray}
which is precisely the desired two point correlation when evaluated at the boundary ($ u=0 $)\cite{Son:2002sd}-\cite{Policastro:2002se}. The natural next step for us would be to solve (\ref{E15}) in order to obtain the precise expression for the function $ \mathcal{F}(u,\omega) $ at the boundary of the $ RNSch_5 $ space time. In order to solve (\ref{E15}), based on the incoming wave boundary condition \cite{Policastro:2002se} we consider the following ansatz namely,
\begin{eqnarray}
f_{x(\omega)} = \left( 1-\frac{u}{u_H}\right)^{\alpha}g_{\omega}(u)\label{E18}
\end{eqnarray} 
where $ g_{\omega}(u) $ is some function that is regular at horizon ($ u=1 $). Our next task would be to find an exact expression for the coefficient $ \alpha $ consistent with the incoming wave boundary condition. Substituting (\ref{E18}) into (\ref{E15}) and considering the near horizon ($ u \sim u_H $) limit we arrive at the following quadratic equation for $ \alpha $ namely,
\begin{eqnarray}
\alpha^{2} + \frac{\omega^{2}u_H^{2}}{4(2-\mathcal{Q}^{2})^{2}}=0.\label{E19}
\end{eqnarray}
From the above equation (\ref{E19}) one can immediately read off $ \alpha $ such that the ansatz (\ref{E18}) is consistent with the incoming wave boundary condition namely\footnote{Note that in order for $ \alpha $ to be dimensionless $ \omega $ should be assigned with the inverse dimension of length namely, $ [\omega] = L^{-1}$.},
\begin{eqnarray}
\alpha =- \frac{i \omega}{4 \pi T}.
\end{eqnarray} 

As a next step we substitute (\ref{E18}) into (\ref{E15}) which finally yields,
\begin{eqnarray}
g_{\omega}'' + \left[\frac{i\omega}{2 \pi u_H  T(1-u/u_H)}+\frac{b'}{b}\right]g_{\omega}' +\frac{i\omega}{4\pi u_H T(1-u/u_H)}\left[\frac{b'}{b}+\frac{1}{u_H(1-u/u_H)} \right]g_{\omega}\nonumber\\
 - \frac{\omega^{2}}{16}\left[ \frac{1}{\pi^{2} u_H^{2}T^{2}(1-u/u_H)^{2}}+\frac{16 c}{b}\right]g_{\omega}=0.\label{E21}  
\end{eqnarray}

The next natural step would be to solve (\ref{E21}) perturbatively in the frequency $ \omega $. To do that we consider the following perturbative expansion of the function $ g_{\omega}(u) $ namely,
\begin{eqnarray}
g_{\omega}(u)=g_{\omega}^{(0)}+ i (\omega/ T) g_{\omega}^{(1)} + \mathcal{O}((\omega/ T)^{2})\label{E22}
\end{eqnarray}
where we have implicitly assumed that $ \omega /T \ll 1 $ which essentially corresponds to the so called hydrodynamic limit. Also remember that we are interested to obtain a solution upto leading order in the frequency ($ \omega $). Substituting (\ref{E22}) into (\ref{E21}) and collecting terms upto leading order in $ \omega $ we finally arrive at the following set of equations namely,
\begin{eqnarray}
g_{\omega}^{''(1)} + \frac{1}{2 \pi u_H (1-u/u_H)}g_{\omega}^{'(0)}+\frac{b'}{b}g_{\omega}^{'(1)} +\frac{1}{4\pi u_H (1-u/u_H)}\left[\frac{b'}{b}+\frac{1}{u_H(1-u/u_H)} \right]g_{\omega}^{(0)}&=&0\nonumber\\
g_{\omega}^{''(0)}+\frac{b'}{b}g_{\omega}^{'(0)}&=&0.\nonumber\\
\label{E23}
\end{eqnarray}

As the above set of equations (\ref{E23}) turn out to be too complicated to be solved exactly in the coordinate ($ u $) therefore like in the previous case we note down the corresponding solutions near the boundary of the $ RNSch_5 $ space time,
\begin{eqnarray}
g_{\omega}^{(1)}&=& \mathcal{C}_{3} \left(-\frac{u u_H}{2}-\frac{u_H^2}{4}\right) e^{-\frac{2 u}{u_H}}-\frac{\mathcal{C}_{6} u}{4 \pi  u_H}+\mathcal{C}_{4}\nonumber\\
g_{\omega}^{(0)}&=& \frac{\mathcal{C}_{5} u^2}{2}+\mathcal{C}_{6}.
\label{E24}
\end{eqnarray}
 A few comments are to be made at this stage regarding the above solution in (\ref{E24}). First of all, note that since the function $ g_{\omega} $ is dimensionful namely, $ [g_{\omega}] = L$ therefore the arbitrary coefficients ($ \mathcal{C}_{i} (i=3,4,5,6) $) must also have dimensions namely, $ [\mathcal{C}_{3}]=[\mathcal{C}_{5}]=L^{-1} $ and $[\mathcal{C}_{4}]=[\mathcal{C}_{6}]=L$. Moreover all of these coefficients are not independent, some of them are rather constrained by the condition, $ f_{a(\omega)}(u)|_{u=0}=L $.
 
Substituting (\ref{E24}) into (\ref{E17}) the DC conductivity finally turns out to be\footnote{Note that we have rescaled the DC conductivity by the volume factor. Therefore strictly speaking we should keep in mind that this is the conductivity per unit volume.},
\begin{eqnarray}
\sigma_{DC}=\frac{L}{8 \pi u_H^{2}g_{SG}^{2}T}\left[\pi u_H^{4}\mathcal{C}_{3}\mathcal{C}_{5}-4 \pi u_H^{2}(\mathcal{C}_{4}\mathcal{C}_{5}-\mathcal{C}_{3}\mathcal{C}_{6})+\mathcal{C}^{2}_{6}\right] .\label{E25}
\end{eqnarray}

It is now in fact a quite trivial exercise to figure out that the DC conductivity (from the dimensional arguments) scales with the temperature as $\sigma_{DC}\sim T^{-1} $. This fact is consistent with the earlier observations in \cite{Ammon:2010eq} and which is essentially the gravity reflection of the fact that we are working with a scale invariant theory at $ z=2 $ fixed point. At this point it is also customary to mention that taking the limit $ \mathcal{Q}\rightarrow 0 $ in the above expression (\ref{E25}) one should be able to recover the DC conductivity corresponding to the uncharged Schr{\"o}dinger black brane. 
Finally, using (\ref{E9}) and (\ref{E25}) the diffusion constant corresponding to $ RNSch_5 $ black brane turns out to be,
\begin{eqnarray}
\mathcal{D}&=&\sigma_{DC}/\chi = \frac{e^{2}\Xi(2\mathcal{C}_2 \left(\beta ^4 L^8 \left(\mathcal{Q}^2+1\right)+3u_H^{4}\right)^2-9\mathcal{C}_1 u_H^{8}\beta ^4 L^8)}{8 \pi u_H^{2} L^{4}\beta ^2\mathcal{C}_1  T\left(\beta ^4 L^8
 \left(\mathcal{Q}^2+1\right)+3 u_H^{4}\right)^2}\nonumber\\
\Xi &=& \pi u_H^{4}\mathcal{C}_{3}\mathcal{C}_{5}-4 \pi u_H^{2}(\mathcal{C}_{4}\mathcal{C}_{5}-\mathcal{C}_{3}\mathcal{C}_{6})+\mathcal{C}^{2}_{6}.
\label{E26}
\end{eqnarray}

Eq.(\ref{E26}) essentially encodes one of the major findings of this paper namely the charge diffusion constant for non relativistic hyperscaling preserving CFTs in the large $ N $ limit where the field theory is essentially at strong coupling. This result although differs non trivially from the pure AdS case, is remarkably simple in the sense that it is independent of the t'Hooft coupling like one finds in the pure AdS case \cite{Kovtun:2008kx}. 

Let us now try to extract the physics out of (\ref{E26}). To start with we note that since all the coefficients $ \mathcal{C}_{i} $ are dimensionful therefore from the above expression (\ref{E26}) it is in fact quite easy to note that the charge diffusion coefficient ($ \mathcal{D} $) has the cubic dimension of length\footnote{This is essentially the consequence of the fact that the charge susceptibility ($ \chi $) goes with temperature as $\chi \sim T^{2} $, whereas on the other hand the conductivity scales as $\sigma_{DC}\sim T^{-1}  $. Therefore the ratio must be a dimensionful entity.} namely, $[\mathcal{D}]=L^{3}$. The take home message from the above analysis is that unlike the pure AdS \cite{Kovtun:2008kx} the charge diffusion coefficient (\ref{E26}) for the $ RNSch_5 $ black branes scales quite non trivially with the temperature namely, $ \mathcal{D} \sim T^{-3} $. This is indeed an interesting observation in the sense that for non relativistic branes the $ \sigma_{DC}/\chi $ ratio is more suppressed at large values of the temperature compared to that of the pure AdS case \cite{Kovtun:2008kx}-\cite{Kovtun:2003wp}, \cite{Policastro:2002se}. The above observation therefore confirms the non universality of the $ \sigma_{DC}/\chi $ ratio for non relativistic branes preserving the hyperscaling and also distinctly categorize them from that of the pure AdS black branes as long as the charge transport phenomena is concerned. 
\section{Summary and final remarks}
In this section we summarize the key findings of our analysis along with mentioning some of its interesting future extensions. The focus of the entire analysis was to explore the (non)universality of the $ \sigma_{DC}/\chi $ ratio in context of non relativistic holography. In our analysis, based on the prescription of \cite{Son:2002sd}-\cite{Policastro:2002se} we have computed the retarded two point correlator for $ z=2 $ non relativistic CFT at strong coupling whose dual gravitational counterpart consists of hyperscaling preserving charged Schr{\"o}dinger black branes \cite{Cremonesi:2009gy} in ($ 4+1 $) dimensions. From our analysis we note that, unlike the pure AdS case \cite{Kovtun:2008kx}, the diffusion coefficient ($ \mathcal{D} $) for charged Schr{\"o}dinger black branes \cite{Cremonesi:2009gy} scales quite non trivially with the temperature namely, $ \mathcal{D}\sim T^{-3} $. This essentially suggests a possible violation of the universality of $ \sigma_{DC}/\chi $ ratio for non relativistic CFTs at strong coupling and could be considered as the gravity reflection of the fact that we are basically working with a $ z=2 $ scale invariant theory at finite temperature.

One crucial fact about our analysis is that in the present analysis we have restricted ourselves at the two derivative level in the action (\ref{E1}). As a natural consequence of this fact the two point correlation function between the $ U(1) $ operators that we have computed essentially stands for the two point correlation within the framework of a non relativistic CFT in the large $ N $ limit. As a part of the future investigations one might go beyond the two derivative order and try to see the effects of adding $ 1/N $ corrections on the two point correlation function of the non relativistic CFT. Note that in the language of gauge/gravity duality the effect of adding $ 1/N $ corrections on the dual CFT side could be incorporated in the form of higher derivative corrections in the bulk theory. Therefore in order to explore such non trivial effects in the framework of non relativistic holography one might replace the usual Maxwell action (\ref{E1}) by the its Weyl corrected form and this could be considered as an extension of the earlier analysis \cite{Ritz:2008kh} to non relativistic branes.

{\bf {Acknowledgements :}}
 The author would like to acknowledge the financial support from CHEP, Indian Institute of Science, Bangalore.\\



\begin{thebibliography}{99}
\bibitem{Taylor:2008tg} 
  M.~Taylor,
  ``Non-relativistic holography,''
  arXiv:0812.0530 [hep-th].
  \bibitem{Son:2008ye} 
  D.~T.~Son,
  ``Toward an AdS/cold atoms correspondence: A Geometric realization of the Schrodinger symmetry,''
  Phys.\ Rev.\ D {\bf 78}, 046003 (2008)
  [arXiv:0804.3972 [hep-th]].

\bibitem{Balasubramanian:2008dm} 
  K.~Balasubramanian and J.~McGreevy,
  ``Gravity duals for non-relativistic CFTs,''
  Phys.\ Rev.\ Lett.\  {\bf 101}, 061601 (2008)
  [arXiv:0804.4053 [hep-th]].
  
  \bibitem{Adams:2008wt} 
  A.~Adams, K.~Balasubramanian and J.~McGreevy,
  ``Hot Spacetimes for Cold Atoms,''
  JHEP {\bf 0811}, 059 (2008)
  [arXiv:0807.1111 [hep-th]].
  
  \bibitem{Maldacena:2008wh} 
  J.~Maldacena, D.~Martelli and Y.~Tachikawa,
  ``Comments on string theory backgrounds with non-relativistic conformal symmetry,''
  JHEP {\bf 0810}, 072 (2008)
  [arXiv:0807.1100 [hep-th]].
  
  \bibitem{Herzog:2008wg} 
  C.~P.~Herzog, M.~Rangamani and S.~F.~Ross,
  ``Heating up Galilean holography,''
  JHEP {\bf 0811}, 080 (2008)
  [arXiv:0807.1099 [hep-th]].
  
  \bibitem{Goldberger:2008vg} 
  W.~D.~Goldberger,
  ``AdS/CFT duality for non-relativistic field theory,''
  JHEP {\bf 0903}, 069 (2009)
  [arXiv:0806.2867 [hep-th]].
  
  \bibitem{Barbon:2008bg} 
  J.~L.~F.~Barbon and C.~A.~Fuertes,
  ``On the spectrum of nonrelativistic AdS/CFT,''
  JHEP {\bf 0809}, 030 (2008)
  [arXiv:0806.3244 [hep-th]].
  
  \bibitem{Kovtun:2008qy} 
  P.~Kovtun and D.~Nickel,
  ``Black holes and non-relativistic quantum systems,''
  Phys.\ Rev.\ Lett.\  {\bf 102}, 011602 (2009)
  [arXiv:0809.2020 [hep-th]].
  
  \bibitem{Duval:2008jg} 
  C.~Duval, M.~Hassaine and P.~A.~Horvathy,
  ``The Geometry of Schrodinger symmetry in gravity background/non-relativistic CFT,''
  Annals Phys.\  {\bf 324}, 1158 (2009)
  [arXiv:0809.3128 [hep-th]].
  
  \bibitem{Ammon:2010eq} 
  M.~Ammon, C.~Hoyos, A.~O'Bannon and J.~M.~S.~Wu,
  ``Holographic Flavor Transport in Schrodinger Spacetime,''
  JHEP {\bf 1006}, 012 (2010)
  [arXiv:1003.5913 [hep-th]].
  
  \bibitem{Kim:2012nb} 
  B.~S.~Kim,
  ``Schr\'odinger Holography with and without Hyperscaling Violation,''
  JHEP {\bf 1206}, 116 (2012)
  [arXiv:1202.6062 [hep-th]].
  
  \bibitem{Adams:2008zk} 
  A.~Adams, A.~Maloney, A.~Sinha and S.~E.~Vazquez,
  ``1/N Effects in Non-Relativistic Gauge-Gravity Duality,''
  JHEP {\bf 0903}, 097 (2009)
  [arXiv:0812.0166 [hep-th]].
  
  \bibitem{Kim:2010zq} 
  B.~S.~Kim, E.~Kiritsis and C.~Panagopoulos,
  ``Holographic quantum criticality and strange metal transport,''
  New J.\ Phys.\  {\bf 14}, 043045 (2012)
  [arXiv:1012.3464 [cond-mat.str-el]].
  \bibitem{Kim:2010tf} 
  B.~S.~Kim and D.~Yamada,
  ``Properties of Schroedinger Black Holes from AdS Space,''
  JHEP {\bf 1107}, 120 (2011)
  [arXiv:1008.3286 [hep-th]].
  
  \bibitem{Mazzucato:2008tr} 
  L.~Mazzucato, Y.~Oz and S.~Theisen,
  ``Non-relativistic Branes,''
  JHEP {\bf 0904}, 073 (2009)
  [arXiv:0810.3673 [hep-th]].
  
  \bibitem{Adams:2011kb} 
  A.~Adams and J.~Wang,
  ``Towards a Non-Relativistic Holographic Superfluid,''
  New J.\ Phys.\  {\bf 13}, 115008 (2011)
  [arXiv:1103.3472 [hep-th]].
  
  \bibitem{Hartong:2010ec} 
  J.~Hartong and B.~Rollier,
  ``Asymptotically Schroedinger Space-Times: TsT Transformations and Thermodynamics,''
  JHEP {\bf 1101}, 084 (2011)
  [arXiv:1009.4997 [hep-th]].
  
  \bibitem{Karch:2014mba} 
  A.~Karch,
  ``Conductivities for Hyperscaling Violating Geometries,''
  JHEP {\bf 1406}, 140 (2014)
  [arXiv:1405.2926 [hep-th]].
  
  \bibitem{Edalati:2013tma} 
  M.~Edalati and J.~F.~Pedraza,
  ``Aspects of Current Correlators in Holographic Theories with Hyperscaling Violation,''
  Phys.\ Rev.\ D {\bf 88}, 086004 (2013)
  [arXiv:1307.0808 [hep-th]].
  
   \bibitem{Kim:2012pd} 
  B.~S.~Kim,
  ``Hyperscaling violation : a unified frame for effective holographic theories,''
  JHEP {\bf 1211}, 061 (2012)
  [arXiv:1210.0540 [hep-th]].
  
   \bibitem{Gath:2012pg} 
  J.~Gath, J.~Hartong, R.~Monteiro and N.~A.~Obers,
  ``Holographic Models for Theories with Hyperscaling Violation,''
  JHEP {\bf 1304}, 159 (2013)
  [arXiv:1212.3263 [hep-th]].
  
  \bibitem{Momeni:2012tw} 
  D.~Momeni, R.~Myrzakulov, L.~Sebastiani and M.~R.~Setare,
  ``Analytical holographic superconductors in $AdS_N$ topological Lifshitz black holes,''
  arXiv:1210.7965 [hep-th].
  
  \bibitem{Guica:2010sw} 
  M.~Guica, K.~Skenderis, M.~Taylor and B.~C.~van Rees,
  ``Holography for Schrodinger backgrounds,''
  JHEP {\bf 1102}, 056 (2011)
  [arXiv:1008.1991 [hep-th]].
  
  \bibitem{Fuertes:2009ex} 
  C.~A.~Fuertes and S.~Moroz,
  ``Correlation functions in the non-relativistic AdS/CFT correspondence,''
  Phys.\ Rev.\ D {\bf 79}, 106004 (2009)
  [arXiv:0903.1844 [hep-th]].
  
  \bibitem{Imeroni:2009cs} 
  E.~Imeroni and A.~Sinha,
  ``Non-relativistic metrics with extremal limits,''
  JHEP {\bf 0909}, 096 (2009)
  [arXiv:0907.1892 [hep-th]].
  
  \bibitem{Volovich:2009yh} 
  A.~Volovich and C.~Wen,
  ``Correlation Functions in Non-Relativistic Holography,''
  JHEP {\bf 0905}, 087 (2009)
  [arXiv:0903.2455 [hep-th]].
  
  \bibitem{Rangamani:2008gi} 
  M.~Rangamani, S.~F.~Ross, D.~T.~Son and E.~G.~Thompson,
  ``Conformal non-relativistic hydrodynamics from gravity,''
  JHEP {\bf 0901}, 075 (2009)
  [arXiv:0811.2049 [hep-th]].
  
  \bibitem{Compere:2009qm} 
  G.~Compere, S.~de Buyl, S.~Detournay and K.~Yoshida,
  ``Asymptotic symmetries of Schrodinger spacetimes,''
  JHEP {\bf 0910}, 032 (2009)
  [arXiv:0908.1402 [hep-th]].
  
  \bibitem{Kraus:2011pf} 
  P.~Kraus and E.~Perlmutter,
  ``Universality and exactness of Schrodinger geometries in string and M-theory,''
  JHEP {\bf 1105}, 045 (2011)
  [arXiv:1102.1727 [hep-th]].
  
  \bibitem{Sadeghi:2012vv} 
  J.~Sadeghi, B.~Pourhasan and F.~Pourasadollah,
  ``Thermodynamics of Schrödinger black holes with hyperscaling violation,''
  Phys.\ Lett.\ B {\bf 720}, 244 (2013)
  [arXiv:1209.1874 [hep-th]].
  
  \bibitem{Perlmutter:2012he} 
  E.~Perlmutter,
  ``Hyperscaling violation from supergravity,''
  JHEP {\bf 1206}, 165 (2012)
  [arXiv:1205.0242 [hep-th]].
  
  \bibitem{Alishahiha:2009nm} 
  M.~Alishahiha, R.~Fareghbal, A.~E.~Mosaffa and S.~Rouhani,
  ``Asymptotic symmetry of geometries with Schrodinger isometry,''
  Phys.\ Lett.\ B {\bf 675}, 133 (2009)
  [arXiv:0902.3916 [hep-th]].
  
  \bibitem{Yamada:2008if} 
  D.~Yamada,
  ``Thermodynamics of Black Holes in Schrodinger Space,''
  Class.\ Quant.\ Grav.\  {\bf 26}, 075006 (2009)
  [arXiv:0809.4928 [hep-th]].
  
  \bibitem{Wang:2013tv} 
  J.~Wang,
  ``Schrodinger Fermi Liquids,''
  Phys.\ Rev.\ D {\bf 89}, 046008 (2014)
  [arXiv:1301.1986 [hep-th]].
  
  \bibitem{Kovtun:2008kx} 
  P.~Kovtun and A.~Ritz,
  ``Universal conductivity and central charges,''
  Phys.\ Rev.\ D {\bf 78}, 066009 (2008)
  [arXiv:0806.0110 [hep-th]].
  
  \bibitem{Kovtun:2003wp} 
  P.~Kovtun, D.~T.~Son and A.~O.~Starinets,
  ``Holography and hydrodynamics: Diffusion on stretched horizons,''
  JHEP {\bf 0310}, 064 (2003)
  [hep-th/0309213].
  
  \bibitem{Son:2002sd} 
  D.~T.~Son and A.~O.~Starinets,
  ``Minkowski space correlators in AdS / CFT correspondence: Recipe and applications,''
  JHEP {\bf 0209}, 042 (2002)
  [hep-th/0205051].

  \bibitem{Policastro:2002se} 
  G.~Policastro, D.~T.~Son and A.~O.~Starinets,
  ``From AdS / CFT correspondence to hydrodynamics,''
  JHEP {\bf 0209}, 043 (2002)
  [hep-th/0205052].
  
  \bibitem{Mas:2008qs} 
  J.~Mas, J.~P.~Shock and J.~Tarrio,
  ``A Note on conductivity and charge diffusion in holographic flavour systems,''
  JHEP {\bf 0901}, 025 (2009)
  [arXiv:0811.1750 [hep-th]].
  
  \bibitem{Herzog:2002fn} 
  C.~P.~Herzog,
  ``The Hydrodynamics of M theory,''
  JHEP {\bf 0212}, 026 (2002)
  [hep-th/0210126].
  
  \bibitem{Adams:2009dm} 
  A.~Adams, C.~M.~Brown, O.~DeWolfe and C.~Rosen,
  ``Charged Schrodinger Black Holes,''
  Phys.\ Rev.\ D {\bf 80}, 125018 (2009)
  [arXiv:0907.1920 [hep-th]].
  
  \bibitem{Cremonesi:2009gy} 
  S.~Cremonesi, D.~Melnikov and Y.~Oz,
  ``Stability of Asymptotically Schro dinger RN Black Hole and Superconductivity,''
  JHEP {\bf 1004}, 048 (2010)
  [arXiv:0911.3806 [hep-th]].
  
  \bibitem{Pang:2009wa} 
  D.~W.~Pang,
  ``Conductivity and Diffusion Constant in Lifshitz Backgrounds,''
  JHEP {\bf 1001}, 120 (2010)
  [arXiv:0912.2403 [hep-th]].
  
  \bibitem{Lemos:2011gy} 
  J.~P.~S.~Lemos and D.~W.~Pang,
  ``Holographic charge transport in Lifshitz black hole backgrounds,''
  JHEP {\bf 1106}, 122 (2011)
  [arXiv:1106.2291 [hep-th]].
  
   \bibitem{Pang:2009pd} 
  D.~W.~Pang,
  ``On Charged Lifshitz Black Holes,''
  JHEP {\bf 1001}, 116 (2010)
  [arXiv:0911.2777 [hep-th]].
  
  \bibitem{Brynjolfsson:2009ct} 
  E.~J.~Brynjolfsson, U.~H.~Danielsson, L.~Thorlacius and T.~Zingg,
  ``Holographic Superconductors with Lifshitz Scaling,''
  J.\ Phys.\ A {\bf 43}, 065401 (2010)
  [arXiv:0908.2611 [hep-th]].
  
  \bibitem{Gubser:2009qm} 
  S.~S.~Gubser, C.~P.~Herzog, S.~S.~Pufu and T.~Tesileanu,
  ``Superconductors from Superstrings,''
  Phys.\ Rev.\ Lett.\  {\bf 103}, 141601 (2009)
  [arXiv:0907.3510 [hep-th]].
  
  \bibitem{Ritz:2008kh} 
  A.~Ritz and J.~Ward,
  ``Weyl corrections to holographic conductivity,''
  Phys.\ Rev.\ D {\bf 79}, 066003 (2009)
  [arXiv:0811.4195 [hep-th]].
  

\end{thebibliography}
\end{document}